\documentclass[11pt,superscriptaddress,aps,prd,preprint]{revtex4}
\usepackage{amsmath}

\makeatletter
\usepackage[T1]{fontenc}
\usepackage{amsmath}
\usepackage{amssymb}
\usepackage{graphicx}

\usepackage{slashed}

\newcommand{\bea}{\begin{eqnarray}}
\newcommand{\eea}{\end{eqnarray}}

\begin{document}

\title{On Stefan-Boltzmann law and the Casimir effect  at finite temperature in the Schwarzschild spacetime}

\author{A. F. Santos}\email[]{alesandroferreira@fisica.ufmt.br}
\affiliation{Instituto de F\'{\i}sica, Universidade Federal de Mato Grosso,\\
78060-900, Cuiab\'{a}, Mato Grosso, Brazil}

\author{S. C. Ulhoa}\email[]{sc.ulhoa@gmail.com}
\affiliation{International Center of Physics, Instituto de F\'isica, Universidade de Bras\'ilia, 70910-900, Bras\'ilia, DF, Brazil.} \affiliation{Canadian Quantum Research Center, 204-3002, 32 Ave, Vernon, BC V1T 2L7,  Canada.}

\author{Faqir C. Khanna\footnote{Professor Emeritus - Physics Department, Theoretical Physics Institute, University of Alberta\\
Edmonton, Alberta, Canada}}\email[]{khannaf@uvic.ca}
\affiliation{Department of Physics and Astronomy, University of Victoria,\\
3800 Finnerty Road Victoria, BC, Canada}

\begin{abstract}

This paper deals with quantum field theory in curved space-time using the Thermo Field Dynamics. The scalar field is coupled to the Schwarzschild  space
time and then thermalised. The Stefan-Boltzmann law is established at finite temperature and the entropy of the field is calculated. Then the Casimir energy and pressure are obtained at zero and finite temperature.

\end{abstract}

\maketitle

\section{Introduction}

Quantum Field Theory (QFT) has been a very successful theory in physics \cite{general}. Together with general relativity it constitutes a great pillar of modern physics.  Nevertheless there  is no widely recognized theory that unifies a diverse areas of physics even though there are numerous attempts to establish a quantum theory of gravitation \cite{ulhoa, loop, strings}. There are attempts to unify through a Quantum Field Theory in curved space-time. This strand has been widely investigated in understanding the fundamental characteristics of black holes.  Particuarly the  no-hair theorem applied to the Kerr black hole has been considered \cite{Herdeiro} in which a scalar field was coupled to the black hole physics and the field equations is solved by using numerical technique.

On the other hand QFT in curved space leads to the well known Hawking effect, that is, the gravitational field creates particles near the event horizon of a black hole, which in turn emits a spectrum of thermal radiation.  This effect is not experimentally verified but systems analogous to black holes in condensed matter reveal that this effect may be real \cite{quasiBH}.  It is important to note that the thermalisation of a field in a curved space-time is not unique, but its importance is fundamental from the point of view of physical reality and everything happens at finite temperature.  There are different ways to introduce temperature under such conditions.  The most widespread form for this purpose is to associate time with temperature by a Wick rotation \cite{Matsubara}.  However this procedure has an undesirable characteristic under certain conditions, the information about the field dynamics, that is due to the temporal coordinate is lost.  However in this article, a different approach is based on Thermo Field Dynamics (TFD), temperature is introduced by doubling the Fock space \cite{Umezawa1, Umezawa2, Umezawa22, Khanna1, Khanna2}.  The main advantage of TFD is that the duplicate space topology allows a study of the Stefan-Boltzmann law and the Casimir effect. Both effects are manifestations of the invariance of Bogoliubov transformation in the Fock space. The Casimir effect is well known in the literature \cite{casimir}, as well as its interaction with the scalar field \cite{scalarcasimir}, even at finite temperature \cite{scalarcasimirtfd}. On the other hand, the investigation of this physical system in different geometries different from the flat space has revealed that Casimir force plays an important role \cite{casimirforce}. Over the years TFD has been applied in various fields such as the scalar field and Dirac field to calculate both Stefan-Boltzmann law and Casimir effect at finite temperature \cite{tfd1, tfd2, tfd3, tfd4}. In this sense, we propose to advance the investigation of TFD in curved spaces.  Schwarzschild space-time is chosen for this purpose fundamentally for two reasons.  The first is of an experimental nature, that is, given the advance of experimental techniques in recent years, it is hoped that some measure of the Casimir effect in Schwarschild space can be verified soon, for instance in the vicinity of the Sun. The second reason is  that gravitational thermodynamics was initially developed in such a curved space.  Thus, from a theoretical point of view, we hope that thermalization by TFD will lead to an understanding of the entropy of this space, different from that normally accepted.

This article is divided as follows. In section II, the scalar field is described in a curved space. The energy-momentum tensor is considered in Schwarzschild space-time. In section III, TFD formalism is introduced. And the generalized Bogoliubov transformation is defined. In section IV, Stefan-Boltzmann law in Schwarzschild space-time is defined and the entropy of the scalar field is calculated. In section V, the Casimir effect at zero temperature in the curved space is established. In section VI, the Casimir effect in Schwarzschild space is calculated at finite temperature. Finally section VII, some brief conclusions are presented.

\section{Gravity coupled to a scalar field}

The action of the mass zero scalar field in curved background is given by
\bea
S=\frac{1}{2}\int d^4x\sqrt{-g}\left(g^{\mu\nu}\partial_\mu\phi(x)\partial_\nu\phi(x)-\xi R\phi(x)^2\right),
\eea
where $g$ is the metric determinant and $\xi$ is the coupling parameter for the scalar curvature, $R$. Then the field equation in curved space-time is
\bea
\left(\partial_\mu\partial^\mu+\xi R\right)\phi(x)=0.
\eea

The energy-momentum tensor, which is defined as
\bea
T_{\gamma\rho}=-\frac{2}{\sqrt{-g}}\frac{\delta S}{\delta g^{\gamma\rho}},
\eea
for a scalar field that leads to   
\bea
T_{\gamma\rho}&=&\frac{1}{2}g_{\gamma\rho}\partial^\mu\phi(x)\partial_\mu\phi(x)-\partial_\gamma\phi(x)\partial_\rho\phi(x)+\xi\left(R_{\gamma\rho}-\frac{1}{2}g_{\gamma\rho}R+g_{\gamma\rho}\Box-\partial_\gamma\partial_\rho\right)\phi(x)^2.
\eea

Now the energy-momentum tensor is used to calculate the Stefan-Boltzmann law and the Casimir effect at finite temperature. In order to  avoid divergences, the energy-momentum tensor is written at different space-time points leading to
\bea
T_{\gamma\rho}(x)&=&\lim_{x'\rightarrow x}\tau\Biggl[\frac{1}{2}g_{\gamma\rho}\partial^\mu\phi(x)\partial_\mu\phi(x')-\partial_\gamma\phi(x)\partial_\beta\phi(x')\nonumber\\&+&\xi\left(R_{\gamma\rho}-\frac{1}{2}g_{\gamma\rho}R+g_{\gamma\rho}\Box-\partial_\gamma\partial_\rho\right)\phi(x)\phi(x')\Biggl],\label{EMT}
\eea
where $\tau$ is the time ordering operator. 

Considering the canonical quantization, the commutation relation is
\bea
[\phi(x),\partial'^\mu\phi(x')]=in_0^\mu\delta({\vec{x}-\vec{x'}}),
\eea
where $n_0^\mu=(1,0,0,0)$ is a time-like vector and 
\bea
\partial^\rho\theta(x_0-x_0')=n_0^\rho\,\delta(x_0-x_0'),
\eea
where $\theta(x_0-x_0')$ is the step function. The energy-momentum tensor becomes
\bea
T_{\gamma\rho}(x)&=&\lim_{x'\rightarrow x}\Bigl\{\Gamma_{\gamma\rho}\tau\left[\phi(x)\phi(x')\right]-I_{\gamma\rho}\delta(x-x')\Bigl\},
\eea
with
\bea
\Gamma_{\gamma\rho}=\frac{1}{2}g_{\gamma\rho}\partial^\mu\partial_\mu-\partial_\gamma\partial_\rho+\xi\left(R_{\gamma\rho}-\frac{1}{2}g_{\gamma\rho}R+g_{\gamma\rho}\Box-\partial_\gamma\partial_\rho\right)\label{Gamma}
\eea
and
\bea
I_{\gamma\rho}=-\frac{i}{2}g_{\gamma\rho}n_0^\mu\,n_{0\mu}+in_{0\gamma}n_{0\rho}.
\eea
In order to calculate the Stefan-Boltzmann law and the Casimir effect associated with the scalar field in the curved space-time, the average momentum of the energy-momentum tensor is determined. Then
\bea
\left\langle T_{\gamma\rho}(x)\right\rangle=\lim_{x'\rightarrow x}\Bigl\{i\Gamma_{\gamma\rho}G_0(x-x')-I_{\gamma\rho}\delta(x-x')\Bigl\},
\eea
where
\bea
iG_0(x-x')=\left\langle 0\left|\tau[\phi(x)\phi(x')]\right| 0 \right\rangle,
\eea
is the scalar field propagator.

\subsection{The Schwarzschild metric}

The Schwarzschild metric as a cosmological background is considered. This metric provides the solution to the Einstein field equations that describe the gravitational field outside a spherical mass. In spherical coordinates $\{t, r, \theta, \phi\}$ this solution is given by
\bea
ds^2=\left(1-\frac{2M}{r}\right)dt^2-\left(1-\frac{2M}{r}\right)^{-1}dr^2+r^2 d\Omega^2,
\eea
where $M$ is mass of the black hole and
\bea
d\Omega^2=d\theta^2+\sin^2\theta d\phi^2.
\eea
Since the interest is in the solution outside a spherical body (the  black hole) the Einstein equation in vacuum is given as
\bea
R_{\gamma\rho}=0.
\eea

Using this equation in vacuum, the quantity $\Gamma_{\gamma\rho}$ defined in eq. (\ref{Gamma}) becomes
\bea
\Gamma_{\gamma\rho}=\frac{1}{2}g_{\gamma\rho}\partial^\mu\partial_\mu-\partial_\gamma\partial_\rho+\xi\left(g_{\gamma\rho}\Box-\partial_\gamma\partial_\rho\right).\label{G}
\eea

In order to obtain the Stefan-Boltzmann law and the Casimir effect for the scalar field coupled to gravity, in a context where the Schwarzschild space-time is the geometric background, two components of eq. (\ref{G}) are used
\bea
\Gamma_{00}&=&-\frac{1}{2}\partial_0\partial'_0-\left(\frac{1}{2}+\xi\right)\left(1-\frac{2M}{r}\right)\left[\left(1-\frac{2M}{r}\right)\partial_1\partial'_1+\frac{1}{r^2}\partial_2\partial'_2+\frac{1}{r^2\sin^2\theta}\partial_3\partial'_3\right]\label{G00}
\eea
and
\bea
\Gamma_{11}&=&-\frac{1}{2}\partial_1\partial'_1+\left(\frac{1}{2}+\xi\right)\left(1-\frac{2M}{r}\right)^{-1}\left[-\left(1-\frac{2M}{r}\right)^{-1}\partial_0\partial'_0+\frac{1}{r^2}\partial_2\partial'_2+\frac{1}{r^2\sin^2\theta}\partial_3\partial'_3\right].\label{G11}
\eea

Another ingredient that will be considered in this cosmological background is the temperature. Finite temperature is introduced in order to consider the     cosmological background using the TFD formalism.

\section{Thermo Field Dynamics (TFD) formalism}

Thermal quantum field theory with a thermal vacuum, $|0(\beta) \rangle$, is considered, with $\beta=\frac{1}{k_BT}$ where $T$ is the temperature and $k_B$ is the Boltzmann constant. The main objective is to interpret the statistical average of an arbitrary operator ${\cal O}$, as the expectation value in a thermal vacuum, i.e., $\langle {\cal O} \rangle=\langle 0(\beta)| {\cal O}|0(\beta) \rangle$. This interpretation leads to two fundamental conditions: (i) the Hilbert space ${\cal S}$ is doubled and (ii) the Bogoliubov transformation is used. The doubled space consists of a thermal space that is defined as ${\cal S}_T={\cal S}\otimes \tilde{\cal S}$, where $\tilde{\cal S}$ is the dual (tilde) Hilbert space.  The Bogoliubov transformation introduces temperature effects through a rotation between tilde ($\tilde{\cal S}$) and non-tilde (${\cal S}$) operators. 

For arbitrary operators ${\cal O}$ and $\tilde{\cal O}$ in Hilbert space ${\cal S}$ and tilde space $\tilde{\cal S}$ respectively, the Bogoliubov transformation is given as
\bea
\left( \begin{array}{cc} {\cal O}(k, \alpha)  \\\eta \tilde {\cal O}^\dagger(k,\alpha) \end{array} \right)={\cal B}(\alpha)\left( \begin{array}{cc} {\cal O}(k)  \\ \eta\tilde {\cal O}^\dagger(k) \end{array} \right),
\eea
where $k$ is the 4-momentum and $\eta = -1(+1)$ for bosons (fermions) and  ${\cal B}(\alpha)$ is defined as
\bea
{\cal B}(\alpha)=\left( \begin{array}{cc} u(\alpha) & -v(\alpha) \\
\eta v(\alpha) & u(\alpha) \end{array} \right),
\eea
with $u^2(\alpha)+\eta v^2(\alpha)=1$. The $\alpha$ parameter is assumed to be the compactification parameter defined by $\alpha=(\alpha_0,\alpha_1,\cdots\alpha_{D-1})$, with $D$ being the space-time dimension. The effect of temperature is described by the choice $\alpha_0\equiv\beta$ and $\alpha_1,\cdots\alpha_{D-1}=0$. In general, $\alpha$ may be associated with any physical quantity. The functions $u(\alpha)$ and $v(\alpha)$, are related to the Bose distribution, and are given as
\bea
v^2(\alpha)=(e^{\alpha\omega_k}-1)^{-1}, \quad\quad u^2(\alpha)=1+v^2(\alpha),\label{phdef}
\eea
with $\omega_k=k_0$ being the energy.

The TFD formalism is used to introduce the compactification parameter $\alpha$ in the propa\-gator of the theory. Here the scalar field is considered as an example. The propagator of the scalar field is written as
\bea
G_0^{(AB)}(x-x';\alpha)&=&i\langle 0(\alpha)| \tau[\phi^A(x)\phi^B(x')]| 0(\alpha)\rangle,\nonumber\\
&=&i\int \frac{d^4k}{(2\pi)^4}e^{-ik(x-x')}G_0^{(AB)}(k;\alpha),
\eea
where $A, B=1, 2$ represent the doubled space. Then
\bea
G_0^{(AB)}(k;\alpha)={\cal B}^{-1}(\alpha)G_0^{(AB)}(k){\cal B}(\alpha),
\eea
with
\bea
G_0^{(AB)}(k)=\left( \begin{array}{cc} G_0(k) & 0 \\
0 & \eta G^*_0(k) \end{array} \right),
\eea
and
\bea
G_0(k)=\frac{1}{k^2+i\epsilon},
\eea
being the usual massless scalar field propagator. $G^*_0(k)$ is the conjugate complex of $G_0(k)$.

Then the Green function (the physical component is given for $A=B=1$) is
\bea
G_0^{(11)}(k;\alpha)=G_0(k)+\eta\, v^2(k;\alpha)[G^*_0(k)-G_0(k)],
\eea
where $v^2(k;\alpha)$ is the generalized Bogoliubov transformation \cite{GBT} which is given as
\bea
v^2(k;\alpha)=\sum_{s=1}^d\sum_{\lbrace\sigma_s\rbrace}2^{s-1}\sum_{l_{\sigma_1},...,l_{\sigma_s}=1}^\infty(-\eta)^{s+\sum_{r=1}^sl_{\sigma_r}}\,\exp\left[{-\sum_{j=1}^s\alpha_{\sigma_j} l_{\sigma_j} k^{\sigma_j}}\right],\label{BT}
\eea
with $d$ being the number of compactified dimensions, $\eta=1(-1)$ for fermions (bosons), $\lbrace\sigma_s\rbrace$ denotes the set of all combinations with $s$ elements.

In the next section, the TFD formalism is used to calculate the Stefan-Boltzmann law and the Casimir effect at temperature $T$ for a mass zero scalar field coupled to gravity in a geometric background described by the Schwarzschild spacetime.

In order to calculate these quantities for the scalar field coupled to gravitational field, a field theory on the topology $\Gamma_D^d=(\mathbb{S}^1)^d\times \mathbb{R}^{D-d}$ with $1\leq d \leq D$ is considered. Here $d$ is the number of compactified dimensions. This establishes a formalism such that any set of dimensions of the manifold $\mathbb{R}^{D}$ are compactified, where the circumference of the $nth$ $\mathbb{S}^1$ is specified by $\alpha_n$.

\section{Stefan-Boltzmann law in the Schwarzschild spacetime}

In order to calculate the Stefan-Boltzmann law a physical (renormalized) energy-momentum tensor is calculated. Using the Casimir prescription, the finite energy-momentum tensor has the form
\bea
{\cal T}_{\gamma\rho}(x;\alpha)=\left\langle T_{\gamma\rho}^{(AB)}(x;\alpha)\right\rangle-\left\langle T_{\gamma\rho}^{(AB)}(x)\right\rangle,
\eea
where the duplicate notation of TFD formalism is used with one component dependant on constant ($\alpha$). Then
\bea
{\cal T}_{\gamma\rho}(x;\alpha)=\lim_{x'\rightarrow x}\Bigl\{i\Gamma_{\gamma\rho}\overline{G}_0^{(AB)}(x-x';\alpha)\Bigl\},
\eea
where
\bea
\overline{G}_0^{(AB)}(x-x';\alpha)=G_0^{(AB)}(x-x';\alpha)-G_0^{(AB)}(x-x'),\label{Green}
\eea
and constant $\Gamma_{\gamma\rho}$ is given in  eq.(\ref{Gamma}).

In order to calculate the Stefan-Boltzmann law in the Schwarzschild spacetime, the topology $\Gamma_4^1=\mathbb{S}^1\times\mathbb{R}^{3}$, with $\alpha=(\beta,0,0,0)$ is used. Using the generalized Bogoliubov transformation
\bea
v^2(\beta)=\sum_{l_0=1}^{\infty}e^{-\beta k^0l_0},\label{BT1}
\eea
the Green function becomes
\bea
\overline{G}_0(x-x';\beta)=2\sum_{l_0=1}^{\infty}G_0(x-x'-i\beta l_0n_0),\label{GF1}
\eea
where $n_0=(1,0,0,0)$ and $\overline{G}_0(x-x';\beta)\equiv \overline{G}_0^{(11)}(x-x';\beta)$, that is the physical component. Then the energy-momentum tensor is
\bea
{\cal T}^{(11)}_{\gamma\rho}(x;\beta)&=&2i\lim_{x'\rightarrow x}\Bigl\{\Gamma_{\gamma\rho}\sum_{l_0=1}^{\infty}G_0(x-x'-i\beta l_0n_0)\Bigl\}.
\eea

The energy associated with the scalar field coupled to gravity in a Schwarzschild background is obtained by choosing $\gamma=\rho=0$. Then
\bea
{\cal T}^{(11)}_{00}(x;\beta)&=&2i\lim_{x'\rightarrow x}\Bigl\{\Gamma_{00}\sum_{l_0=1}^{\infty}G_0(x-x'-i\beta l_0n_0)\Bigl\},\label{11}
\eea
where $\Gamma_{00}$ is given in eq. (\ref{G00}) and 
\bea
G_0(x-x'-i\beta l_0n_0)=-\frac{i}{(2\pi)^2}\frac{1}{(x-x'-i\beta l_0n_0)^2},\label{GF}
\eea
with
\bea
(x-x'-i\beta l_0n_0)^2&=&-\left(1-\frac{2M}{r}\right)(t-t'-i\beta l_0)^2+\left(1-\frac{2M}{r}\right)^{-1}(r-r')^2\nonumber\\
&+&r^2(\theta-\theta')^2+r^2\sin^2\theta(\phi-\phi')^2.
\eea

Using these results, the $00$-component of the energy-momentum tensor at finite temperature becomes
\bea
{\cal T}^{(11)}_{00}(T)=\frac{\pi^2(1+\xi)r}{30(r-2M)}T^4,\label{SBL}
\eea
where ${\cal T}^{(11)}_{00}(T)\equiv E(T)$ and the Riemann Zeta function,
\bea
\zeta(4)=\sum_{l_0=1}^\infty\frac{1}{l_0^4}=\frac{\pi^4}{90},\label{zetaf}
\eea
is used. The eq. (\ref{SBL}) is the Stefan-Boltzmann law for the black hole. In the limit $r>>2M$, the Stefan-Boltzmann law becomes
\bea
{\cal T}^{(11)}_{00}(T)=\frac{\pi^2(1+\xi)}{30}T^4.\label{SBL1}
\eea
The Stefan-Boltzmann law for the massless scalar field in a flat spacetime is given by ${\cal T}^{(11)}_{00}(T)$, i.e., eq. (\ref{SBL1}). Then in this limit, the gravitational effect due to the black hole is ignored. 

With the pressure given as $P=\frac{E}{3}$, and the Maxwell relationship given by
\bea
\left(\dfrac{\partial P}{\partial T} \right)_{V}=\left(\dfrac{\partial S}{\partial V} \right)_{T},
\eea
where $S$ is the entropy leads to
\bea
\left(\dfrac{\partial S}{\partial V} \right)_{T}=\frac{4\pi^2(1+\xi)r}{90(r-2M)}T^3.
\eea
Then the entropy associated with the scalar field coupled to gravity in a Schwarzschild spacetime is given as
\bea
S=\frac{4\pi^2(1+\xi)}{90}T^3\int \frac{r}{(r-2M)}r^2\sin \theta dr d\theta d\phi.
\eea
Performing an integration, the gravitational entropy becomes
\bea
S=\frac{64M^3\pi^3(1+\xi)}{45}T^3\left[-\ln\left|\frac{2M}{2M-r}\right|+\frac{R}{2M}+\frac{R^2}{8M^2}\right]\,,
\eea
where $R$ is the radius of the integration hypersurface.  It is important to note that this expression has two noteworthy regions of singularity.  The first is the black hole event horizon, $ r = 2M $.  This singularity is not of concern because it stems from the choice of the coordinate system.  In addition the causal region is outside the event horizon.  The singularity in the $ r \rightarrow \infty $ region arises from integration over the whole spacetime.  It should be noted, however, that the entropy density remains finite even in the limit of a flat spacetime.  Thus its integration with infinite space will also lead to a divergence.  It is worth noting that an entropy expression is linked to both macroscopic and microscopic components.  But in the case of gravitational interaction we have no information on the latter.  Thus the spacetime entropy divergence is an expected result as it would arise from, a priori, an infinite ensemble.

\section{Casimir effect at zero temperature}

In order to calculate the Casimir effect at zero temperature, a theory with topology $\Gamma_4^1=\mathbb{S}^1\times\mathbb{R}^{3}$ is considered. Choosing the $\alpha$ parameter as $\alpha=(0,0,0,i2b)$, where $2b$ corresponds to the length of the circumference $\mathbb{S}^1$. Here the Bogoliubov transformation is
\bea
v^2(b)=\sum_{l_3=1}^{\infty}e^{-i2b k^3l_3}\label{BT2}
\eea
and the Green function is
\bea
\overline{G}_0(x-x';b)=2\sum_{l_3=1}^{\infty}G_0(x-x'-2b l_3n_3)\label{GF2}
\eea
with $n_3=(0,0,0,1)$. The Casimir energy is calculated for the case $\gamma=\rho=0$. Then the energy-momentum tensor becomes
\bea
{\cal T}^{(11)}_{00}(x;b)&=&2i\lim_{x'\rightarrow x}\Bigl\{\Gamma_{00}\sum_{l_3=1}^{\infty}G_0(x-x'-2b l_3z)\Bigl\}.\label{111}
\eea
Using eq. (\ref{G00}) the Casimir energy ($E_c\equiv {\cal T}^{(11)}_{00}(x;b)$) at zero temperature associated to the massless scalar fiedl coupled to gravity in a Schwarzschild spacetime is
\bea
E_c=\frac{(r-2M)^2}{1440\pi^2 r^4 b^4}\left[\pi^4(1+\xi)r(2M-r)-180bM(1+2\xi)\zeta(3)\right],
\eea
where $\zeta(3)$ is the Riemann Zeta function. Far from the event horizon, i.e. $r>>2M$, the Casimir energy has the form
\bea
E_c=-\frac{\pi^2(1+\xi)}{1440b^4}.
\eea
This is the result found for the case of the flat spacetime, where the gravitational effects due to black hole are ignored.

Similarly the Casimir pressure is calculated by taking $\gamma=\rho=1$. Then the energy-momentum tensor becomes
\bea
{\cal T}^{(11)}_{11}(x;b)&=&2i\lim_{x'\rightarrow x}\Bigl\{\Gamma_{11}\sum_{l_3=1}^{\infty}G_0(x-x'-2b l_3z)\Bigl\}.\label{33}
\eea
Using eq. (\ref{G11}) the Casimir pressure ($P_c\equiv{\cal T}^{(11)}_{11}(x;b)$) at zero temperature is
\bea
P_c=-\frac{60bM\zeta(3)-\pi^4(1+\xi)r(2M-r)}{480\pi^2r^2b^4}.
\eea
In the limit $r>> 2M$ the Casimir pressure is
\bea
P_c=-\frac{\pi^2(1+\xi)}{480b^4}.
\eea
It is a well-known result in flat spacetime. These results indicate that the gravitational force due to the black hole changes the Casimir energy and pressure for a scalar field at zero temperature.

\section{Casimir effect at finite temperature}

With the topology $\Gamma_4^2=\mathbb{S}^1\times\mathbb{S}^1\times\mathbb{R}^{2}$ for $\alpha=(\beta,0,0,i2b)$, the Casimir effect at finite temperature is calculated. Then the Bogoliubov transformation becomes
\bea
v^2(\beta,b)=\sum_{l_0=1}^\infty e^{-\beta k^0l_0}+\sum_{l_3=1}^\infty e^{-i2bk^3l_3}+2\sum_{l_0,l_3=1}^\infty e^{-\beta k^0l_0-i2bk^3l_3},\label{BT3}
\eea
where the first two terms are associated with the Stefan-Boltzmann law and the Casimir effect at zero temperature and the third term provides the combined effect of temperature and spatial compactification. Indeed, in this case, two compactifications, the time and the other along the $z$-coordinate are explored. Then the Green function is
\bea
\overline{G}_0(x-x';\beta,b)&=&4\sum_{l_0,l_3=1}^\infty G_0\left(x-x'-i\beta l_0n_0-2bl_3n_3\right).\label{GF3}
\eea

Using the Green function the energy-momentum becomes
\bea
{\cal T}^{(11)}_{\gamma\rho}(\beta, b)&=&4i\lim_{x'\rightarrow x}\Bigl\{\Gamma_{\gamma\rho}\sum_{l_0=1}^{\infty}G_0(x-x'-i\beta l_0n_0-2bl_3n_3)\Bigl\}.
\eea
This leads to the Casimir energy at finite temperature as
\bea
{\cal T}^{(11)}_{00}(\beta, b)&=&-\frac{2(2M-r)}{\pi^2r^2}\sum_{l_0,l_3=1}^\infty\frac{1}{[(2bl_3)^2(2M+r)-(\beta l_0)^2(2M-r)]^3}\nonumber\\
&\times&\Bigl[8(bl_3)^3M(1+2\xi)(4M^2-r^2)-(2bl_3)^2r(2M+r)\bigl((1+\xi)r^2\nonumber\\
&-&6M^2(1+2\xi)\bigl)-2(\beta l_0)^2bM(1+2\xi)l_3(2M-r)(2M+3r)\nonumber\\
&+&(\beta l_0)^2r(2M-r)\left(M^2(2+4\xi)-2(1+\xi)r^2\right)\Bigl].
\eea
The Casimir pressure at finite temperature is
\bea
{\cal T}^{(11)}_{11}(\beta, b)&=&\frac{2}{\pi^2(2M-r)}\sum_{l_0,l_3=1}^\infty\frac{1}{[(2bl_3)^2(2M+r)-(\beta l_0)^2(2M-r)]^3}\nonumber\\
&\times&\Bigl[8(l_0l_3)^3M(r^2-4M^2)+3(2bl_3)^2r(2M+r)\bigl((1+\xi)r^2\nonumber\\
&-&2M^2\bigl)+2(\beta l_0)^2bMl_3(2M-r)(2M+3r)\nonumber\\
&-&(\beta l_0)^2r(2M-r)\left(2M^2-(1+\xi)r^2\right)\Bigl].
\eea
Overall the Casimir effect at finite temperature for the scalar field is modified by the gravitational force, due to the presence of the black hole, and finite temperature.

In the limit $r>> 2M$ the Casimir energy and pressure at finite temperature become
\bea
{\cal T}^{(11)}_{00}(\beta, b)&=&-\frac{2}{\pi^2}\sum_{l_0,l_3=1}^\infty\frac{(1+\xi)[2(\beta l_0)^2-(2bl_3)^2]}{[(2bl_3)^2+(\beta l_0)^2]^3}
\eea
and
\bea
{\cal T}^{(11)}_{11}(\beta, b)&=&-\frac{2}{\pi^2}\sum_{l_0,l_3=1}^\infty\frac{(1+\xi)[(\beta l_0)^2-3(2bl_3)^2]}{[(2bl_3)^2+(\beta l_0)^2]^3}.
\eea
These results represent the Casimir energy and pressure in a flat space-time, if the gravitational effect at finite temperature due to the black hole is ignored.

\section{Conclusions}

The Stefan-Boltzmann law is obtained by coupling a scalar field to the Schwarzschild space-time. The thermalization of the scalar field is obtained by using TFD. Using the temperature dependent energy, it is possible to calculate the first law of thermodynamics. The entropy should not be confused with Hawking entropy of Schwarzschild black hole since the first is obtained using the quantum field theory in curved space-time while the second is a geometric property of the black hole. This implies that calculated entropy refers to a property of the scalar field and is dependent on the integration surface. Here the energy and the pressure of the Casimir effect generated by the scalar field in the curved space-time and finite temperature are presented. It is important to note that the Casimir entropy at finite temperature does exist even though the integration over a spherical hypersurface is quite intricate. It is possible that the entropy leads to no remnant since the energy limit for $ T \rightarrow 0 $ is null.

\section*{Acknowledgments}

This work by A. F. S. is supported by CNPq projects 308611/2017-9 and 430194/2018-8.

\end{document}